\documentclass[letter,english]{article}
\usepackage[latin1]{inputenc}
\usepackage{fancyhdr}
\usepackage{graphicx}
\usepackage{amssymb}

\makeatletter

\providecommand{\tabularnewline}{\\}

 \newcommand{\lyxaddress}[1]{
   \par {\raggedright #1 
   \vspace{1.4em}
   \noindent\par}
 }

\usepackage{graphicx}
\usepackage{epsfig}
\usepackage{amsbsy}
\usepackage{amsmath}
\usepackage{amssymb}
\usepackage{html}
\usepackage{hyperref}
\usepackage[abbr,agsmcite]{harvard}

\usepackage{babel}
\makeatother
\begin{document}

\newcommand{\commute}[2]{\left[#1,#2\right]}

\newcommand{\bra}[1]{\left\langle #1\right|}

\newcommand{\ket}[1]{\left|#1\right\rangle }

\newcommand{\anticommute}[2]{\left\{  #1,#2\right\}  }

\renewcommand{\harvardand}{and}

\title{Quantum computing with spins in solids }

\author{W. A. Coish and Daniel Loss}

\maketitle

\lyxaddress{Department of Physics and Astronomy, University of Basel, Klingelbergstrasse
82, CH-4056 Basel, Switzerland}

\begin{abstract}
The ability to perform high-precision one- and two-qubit operations
is sufficient for universal quantum computation. For the Loss-DiVincenzo
proposal to use single electron spins confined to quantum dots as
qubits, it is therefore sufficient to analyze only single- and coupled
double-dot structures, since the strong Heisenberg exchange coupling
between spins in this proposal falls off exponentially with distance
and long-ranged dipolar coupling mechanisms can be made significantly
weaker. This scalability of the Loss-DiVincenzo design is both a practical
necessity for eventual applications of multi-qubit quantum computing
and a great conceptual advantage, making analysis of the relevant
components relatively transparent and systematic. We review the Loss-DiVincenzo
proposal for quantum-dot-confined electron spin qubits, and survey
the current state of experiment and theory regarding the relevant
single- and double- quantum dots, with a brief look at some related
alternative schemes for quantum computing. 
\end{abstract}


\section{INTRODUCTION}

Recent advances in semiconductor spintronics and,
more specifically, spin-based quantum computing in solid-state systems,
have encouraged significant research efforts in the last years \cite{prinz:1998a,wolf:2001a,awschalom:2002a}.
Much of this research is motivated by pressure on the electronics
industry to maintain Moore's-law growth in systems with components
that are very quickly approaching the nanoscale, where quantum mechanics
becomes important \cite{itrs:2005a}. Additionally, nanoscale devices
provide a unique opportunity to study the fundamental physics of quantum
phenomena in a controllable environment. 

Independent of the particular motivation, if quantum information processing
is to progress beyond basic proof-of-principle experiments, it must
be based on a viable, scalable qubit (a quantum mechanical two-level
system, which can be placed in an arbitrary superposition of its basis
states: $\ket{\psi}=a\ket{0}+b\ket{1}$). The two states of single
electron spins ($\ket{\uparrow}=\ket{0}$ and $\ket{\downarrow}=\ket{1}$),
confined to semiconductor quantum dots (the Loss-DiVincenzo proposal),
are one such qubit \cite{loss:1998a}. These qubits are viable, in
the sense that they make use of fabrication techniques and electrical
control concepts that have been developed over the last five decades
in research laboratories and industry. The secret to scalability in
the Loss-DiVincenzo proposal lies in local gating; this proposal would
implement gating operations through the exchange interaction, which
can be tuned locally with exponential precision, allowing pairs of
neighboring qubits to be coupled and decoupled independently. This
is to be contrasted with proposals that make use of long-ranged interactions
(e.g., dipolar coupling) for which scalability may be called into
question. The local, tunable nature of inter-qubit interactions in
the Loss-DiVincenzo proposal is what makes it possible to consider
first isolated one-qubit (single-quantum-dot), then isolated two-qubit
(double-quantum-dot) systems. Once single- and double- quantum dots
are understood, along with environmental couping mechanisms, a quantum
computation can proceed through a series of one- and two-qubit operations,
without great concern regarding interactions between three, four,
and more qubits. 

There are many other proposals for qubits and associated quantum control
processes. Some examples include various proposals that use superconducting
devices (for reviews, see \cite{makhlin:2001a,burkard:2004a}), proposals
for {}``adiabatic quantum computing'', in which quantum computations
are performed through adiabatic manipulation of coupling constants
in physically realizable Hamiltonians \cite{farhi:2000a,farhi:2001a,wu:2005a}
(which might be used to perform fast quantum simulations of, e.g.,
superconducting pairing models \cite{wu:2002a}), electron spin qubits
encoded in two-spin states \cite{levy:2002a} or many-spin chains
\cite{meier:2003a,meier:2003b} (recent work showing that such spin
chains can be built-up atom-by-atom on a surface \cite{hirjibehedin:2006a}
is a possible first step to implementing such a proposal), cavity-QED
schemes \cite{sleator:1995a,domokos:1995a}, trapped-ion proposals
\cite{cirac:1995a}, etc. Each of these proposals has advantages and
disadvantages. Here we do not compare the relative merits of all proposals,
but instead focus on proposals involving electron spins confined to
quantum dots.

Before a quantum computation can begin, the qubits in a working quantum
computer must be initialized to some state, e.g. $\ket{0}$. These
qubits must be sufficiently isolated from the surrounding environment
to reduce decoherence, there must be some way to perform fast single-
and two-qubit operations in a time scale much less than the qubit
decoherence time, and it must be possible to read out the final state
of the qubits after any quantum computation \cite{divincenzo:2000a}.
In the following sections, we address these issues and others which
are important for the Loss-DiVincenzo proposal. In the process, we
survey some recent work on quantum computing with electron spins in
quantum dots. 

Due to the rapid development of this field, there have been many recent
reviews on quantum-dot quantum computing. Among these numerous reviews,
there has been work that focuses on single-electron charge qubits
in double dots \cite{fujisawa:2006a}, the implementation of single-electron
spin resonance (ESR), and the molecular wavefunctions of coupled double
dots \cite{vanderwiel:2006a}, various proposals for spin-based quantum
computing \cite{cerletti:2005a}, silicon-based proposals for quantum
computing \cite{koiller:2005a}, general quantum computing in the
solid state, including both quantum dots and superconducting systems
\cite{burkard:2004a}, experiments and experimental proposals for
quantum-dot-confined electrons \cite{engel:2004b}, the many coupling
schemes and decoherence mechanisms for quantum-dot spin qubits \cite{hu:2004a},
and optical properties of quantum dots \cite{hohenester:2004a}. In
this review, we analyze the Loss-DiVincenzo proposal from the viewpoint
that this proposal can be decomposed into first single and then double
quantum dots, with a special emphasis on double-dot physics.

This review is organized as follows: in Section \ref{sec:LossDiVincenzoOverview}
we give a brief summary of the Loss-DiVincenzo proposal for quantum
computing. In Section \ref{sec:SingleQDs} we discuss the characterization
and manipulation of spin and charge states of electrons in single
quantum dots. Section \ref{sec:DoubleQDs} contains a description
of double quantum dots that emphasizes the single-electron regime,
which is relevant for quantum-dot quantum computing. In Section \ref{sec:Decoherence}
we survey important decoherence mechanisms for electron spins in single-
and double- quantum dots. In Section \ref{sec:Entanglement} we briefly
review some proposals for the generation and detection of nonlocal
entanglement of electron spins in nanostructures, and in Section \ref{sec:Conclusions}
we conclude with a brief summary of important topics for future study.

\section{\label{sec:LossDiVincenzoOverview}SPINS IN QUANTUM DOTS: AN OVERVIEW
OF THE LOSS-DIVINCENZO PROPOSAL}

\begin{figure}
\begin{center}\includegraphics[%
  scale=0.15]{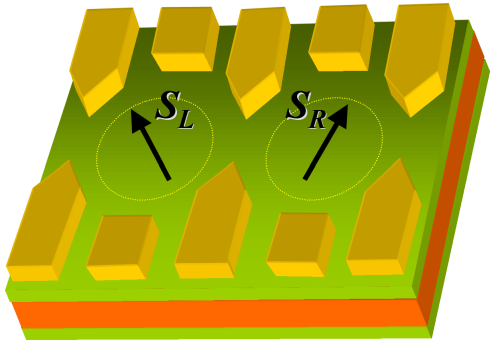}\end{center}

\caption{\label{cap:Double-QD}A double quantum dot. Top-gates are set to
an electrostatic voltage configuration that confines electrons in
the two-dimensional electron gas (2DEG) below to the circular regions
shown. Applying a negative voltage to the back-gate, the dots can
be depleted until they each contain only one single electron, each
with an associated spin-$1/2$ operator $\mathbf{S}_{\mathrm{L}(\mathrm{R})}$
for the electron in the left (right) dot. The $\ket{\uparrow}$ and
$\ket{\downarrow}$ spin-$1/2$ states of each electron provide a
qubit (two-level quantum system). }
\end{figure}
In the original Loss-DiVincenzo proposal, the qubits are stored in
the two spin states of single confined electrons. The considerations
discussed in \cite{loss:1998a} are generally applicable to electrons
confined to any structure (e.g. atoms, molecules, defects, etc.),
although the original proposal focused on applications in gated semiconductor
quantum dots, as shown in Figure \ref{cap:Double-QD}. Voltages applied
to the top gates of such structures provide a confining potential
for electrons in a two-dimensional electron gas (2DEG), below the
surface. A negative voltage applied to a back-gate depletes the 2DEG
locally, allowing the number of electrons in each dot to be reduced
down to one (the single-electron regime). Advances in materials fabrication
and gating techniques have now allowed for the realization of single
electrons in single vertical \cite{tarucha:1996a} and gated lateral
\cite{ciorga:2000a} dots, as well as double dots \cite{elzerman:2003a,hayashi:2003a,petta:2004b}. 

Initialization of all qubits in the quantum computer to the Zeeman
ground state $\ket{\uparrow}=\ket{0}$ could be achieved by allowing
all spins to reach thermal equilibrium at temperature $T$ in the
presence of a strong magnetic field $B$, such that $\left|g\mu_{B}B\right|>k_{B}T$,
with $g$-factor $g<0$, Bohr magneton $\mu_{B}$, and Boltzmann's
constant $k_{B}$ \cite{loss:1998a}. For further initialization schemes,
see Section \ref{sub:SingleSpinControl} below. 

Once the qubits have been initialized to some state, they should remain
in that state until a computation can be executed. In the absence
of environmental coupling, the spins-$1/2$ of single electrons are
intrinsic two-level systems, which cannot {}``leak'' into higher
excited states. Additionally, since electron spins only couple to
charge degrees of freedom indirectly through the spin-orbit (or hyperfine)
interactions, they are relatively immune to fluctuations in the surrounding
electronic environment. 

\begin{figure}
\begin{center}\includegraphics[%
  scale=0.5]{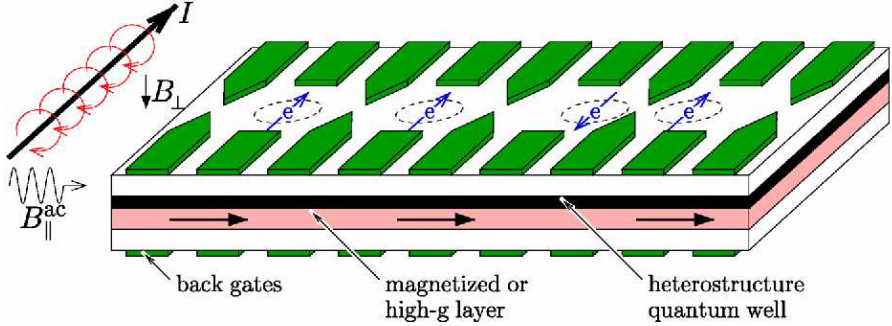}\end{center}

\caption{\label{cap:Series-of-QDs}A series of exchange-coupled electron spins.
Single-qubit operations could be performed in such a structure using
electron spin resonance (ESR), which would require an rf transverse
magnetic field $B_{\parallel}^{\mathrm{ac}}$, and a site-selective
Zeeman splitting $g(x)\mu_{B}B_{\perp}$, which might be achieved
through $g$-factor modulation or magnetic layers. Two-qubit operations
would be performed by bringing two electrons into contact, introducing
a nonzero wavefunction overlap and corresponding exchange coupling
for some time (two electrons on the right). In the idle state, the
electrons can be separated, eliminating the overlap and corresponding
exchange coupling with exponential accuracy (two electrons on the
left). }
\end{figure}

Single-qubit operations in the Loss-DiVincenzo quantum computer could
be carried out by varying the Zeeman splitting on each dot individually
\cite{loss:1998a}. It may be possible to do this through $g$-factor
modulation \cite{salis:2001c}, the inclusion of magnetic layers \cite{myers:2005a}
(see also Figure \ref{cap:Series-of-QDs}), modification of the local
Overhauser field due to hyperfine couplings \cite{burkard:1999a},
or with nearby ferromagnetic dots \cite{loss:1998a}. There are a
number of alternate methods that could be used to perform single-qubit
rotations (see Section \ref{sub:SingleSpinControl}). %

Two-qubit operations would be performed within the Loss-DiVincenzo
proposal by pulsing the exchange coupling between two neighboring
qubit spins {}``on'' to a non-zero value ($J(t)=J_{0}\ne0,\, t\in\left\{ -\tau_{\mathrm{s}}/2\ldots\tau_{\mathrm{s}}/2\right\} $)
for a switching time $\tau_{\mathrm{s}}$, then switching it {}``off''
($J(t)=0,\, t\notin\left\{ -\tau_{\mathrm{s}}/2\ldots\tau_{\mathrm{s}}/2\right\} $).
This switching can be achieved by briefly lowering a center-gate barrier
between neighboring electrons, resulting in an appreciable overlap
of the electron wavefunctions \cite{loss:1998a}, or alternatively,
by pulsing the relative back-gate voltage of neighboring dots \cite{petta:2005a}
(see Section \ref{sub:TwoQubitGates}). Under such an operation (and
in the absence of Zeeman or weaker spin-orbit or dipolar interactions),
the effective two-spin Hamiltonian takes the form of an isotropic
Heisenberg exchange term, given by \cite{loss:1998a,burkard:1999a}

\begin{equation}
H_{\mathrm{ex}}(t)=J(t)\mathrm{\mathbf{S}_{L}\cdot\mathbf{S}_{R}},\label{eq:ExchangeHamiltonian}\end{equation}
 where $\mathbf{S}_{\mathrm{L}(\mathrm{R})}$ is the spin-$1/2$ operator
for the electron in the left (right) dot, as shown in Figure \ref{cap:Double-QD}.
The Hamiltonian $H_{\mathrm{ex}}(t)$ generates the unitary evolution
$U(\phi)=\exp\left[-i\phi\mathrm{\mathbf{S}_{L}\cdot\mathbf{S}_{R}}\right]$,
where $\phi=\int J(t)dt/\hbar$. If the exchange is switched such
that $\phi=\int J(t)dt/\hbar=J_{0}\tau_{\mathrm{s}}/\hbar=\pi$, $U(\phi)$
exchanges the states of the two neighboring spins, i.e.: $U(\pi)\ket{\mathbf{n},\mathbf{n}^{\prime}}=\ket{\mathbf{n}^{\prime},\mathbf{n}}$,
where $\mathbf{n}$ and $\mathbf{n}^{\prime}$ are two arbitrarily
oriented unit vectors and $\ket{\mathbf{n},\mathbf{n}^{\prime}}$
indicates a simultaneous eigenstate of the two operators $\mathbf{S}_{\mathrm{L}}\cdot\mathbf{n}$
and $\mathbf{S}_{\mathrm{R}}\cdot\mathbf{n}^{\prime}$. $U(\pi)$
implements the so-called {\sc swap} operation. If the exchange is
pulsed on for the shorter time $\tau_{\mathrm{s}}/2$, the resulting
operation $U(\pi/2)=\left(U(\pi)\right)^{1/2}$ is known as the {}``square-root-of-{\sc swap}''
($\sqrt{\mbox{{\sc swap}}}$). The $\sqrt{\mbox{{\sc swap}}}$ operation
in combination with arbitrary single-qubit operations is sufficient
for universal quantum computation \cite{barenco:1995b,loss:1998a}.
The $\sqrt{\mbox{{\sc swap}}}$ operation has now been successfully
implemented in experiments involving two electrons confined to two
neighboring quantum dots (as in Figure \ref{cap:Double-QD}) \cite{petta:2005a,laird:2005a}.
Errors during the $\sqrt{\mbox{{\sc swap}}}$ operation have been
investigated due to nonadiabatic transitions to higher orbital states
\cite{schliemann:2001a,requist:2005a}, spin-orbit-interaction \cite{bonesteel:2001a,burkard:2002a,stepanenko:2003a},
and hyperfine coupling to surrounding nuclear spins \cite{petta:2005a,coish:2005a,klauser:2005a,taylor:2006a}.
The isotropic form of the exchange interaction given in Equation (\ref{eq:ExchangeHamiltonian})
is not always valid. In realistic systems, a finite spin-orbit interaction
leads to anisotropic terms which may cause additional errors, but
could also be used to perform universal quantum computing with two-spin
encoded qubits, in the absence of single-spin rotations \cite{bonesteel:2001a,lidar:2002a,stepanenko:2004a,chutia:2006a}
(see also Section \ref{sub:EncodedQubits} below).

In the Loss-DiVincenzo proposal, readout could be performed using
spin-to-charge conversion. This could be accomplished with a {}``spin
filter'' (spin-selective tunneling) to leads or a neighboring dot,
coupled with single-electron charge detection (see also Section \ref{sub:SingleSpinControl},
below).

\section{\label{sec:SingleQDs}SINGLE QUANTUM DOTS}

The fundamental element of information in a quantum computer is the
quantum bit, or qubit. The qubits of the Loss-DiVincenzo proposal
\cite{loss:1998a} are encoded in the two spin states $\ket{\uparrow}$
and $\ket{\downarrow}$ of single electrons confined to quantum dots.

There are many different types of quantum dot that can be manufactured,
each with distinct characteristics. Gated lateral quantum dots (as
shown in Figures \ref{cap:Double-QD} and \ref{cap:Series-of-QDs})
offer the benefit that their shape and size can be controlled to suit
a particular study, and the tunnel coupling between pairs of these
dots can be tuned in a straightforward manner: by raising or lowering
the barrier between the dots. Gated vertical dots \cite{tarucha:1996a}
are created by etching surrounding material to form a pillar structure,
with vertical confinement provided by a double-barrier heterostructure.
Vertical dots allow for the controlled fabrication of quantum dots
with large level spacing, although tunability of the coupling in these
structures is restricted due to the fabrication process. To resolve
this issue, hybrid laterally-coupled vertical double quantum dots
have been manufactured, in which the inter-dot tunnel coupling is
controllable \cite{hatano:2005a}. Self-assembled quantum dots are
yet another type of dot that can be used for quantum information processing.
Self-assembled dots form spontaneously during epitaxial growth due
to a lattice mismatch between the dot and substrate materials. These
dots can be made with very large single-particle level spacings, but
typically form at random locations, which makes controlled coupling
through a tunnel junction difficult. Such dots can, however, potentially
be coupled with optical cavity modes \cite{imamoglu:1999a}, and new
techniques have now allowed the fabrication of cavities with modes
that couple maximally directly at the positions of isolated dots \cite{badolato:2005a}. 

In the rest of this section we focus on lateral quantum dots, as shown
in Figures \ref{cap:Double-QD} and \ref{cap:Series-of-QDs}. After
a brief review of charge and spin control in single quantum dots,
we will address issues specific to double quantum dots in Section
\ref{sec:DoubleQDs}.

\subsection{Charge control: Coulomb blockade}

To ensure a single two-level system is available to be used as a qubit,
it is practical to consider single isolated electron spins (with intrinsic
spin 1/2) confined to single orbital levels. A natural first step
to implementing the Loss-DiVincenzo proposal was therefore to demonstrate
control over the charging of a quantum dot electron-by-electron in
a single gated quantum dot. This is typically done by operating a
quantum dot in the Coulomb-blockade regime, where the energy for the
addition of an electron to the quantum dot is larger than the energy
that can be supplied by electrons in the source or drain leads. In
this case, the charge on the quantum dot is conserved, and no electrons
can tunnel onto or off of the dot. For a general review of Coulomb
blockade phenomena and the characterization of many-electron states
in single quantum dots, see \cite{kouwenhoven:2001a}.

\subsection{\label{sub:SingleSpinControl}Spin control: Initialization, operations,
and readout.}

As mentioned in Section \ref{sec:LossDiVincenzoOverview}, initialization
of all electron spins to the {}``up'' state $\ket{\uparrow}$ could
be achieved by allowing all spins to equilibrate in a strong magnetic
field. Depending on the particular architechture, this may take a
long time or it may be inconvenient to have large magnetic fields
in the region of the apparatus. Initialization could also be achieved
through spin-injection from a ferromagnet, as has been performed in
bulk semiconductors \cite{fiederling:1999a,ohno:1999a}, with a spin-polarized
current from a spin-filter device \cite{prinz:1995a,prinz:1998a,loss:1998a,divincenzo:1999a,recher:2000a},
or by optical pumping \cite{cortez:2002a,shabaev:2003a,gywat:2004a,bracker:2005a},
which has now allowed the preparation of spin states with very high
fidelity, in one case as high as 99.8\% \cite{atature:2006a}.

Single-qubit operations can be performed in the Loss-DiVincenzo proposal
whenever the Zeeman energy of the quantum-dot spins can be tuned locally,
as mentioned in Section \ref{sec:LossDiVincenzoOverview}. Alternative
single-qubit-rotation schemes may require global magnetic field gradients
\cite{wu:2004a,tokura:2006a}, ESR (see Figure \ref{cap:Series-of-QDs})
or, in the presence of spin-orbit interaction, electric-dipole spin
resonance (EDSR) techniques. EDSR has been analyzed in great detail
for two-dimensional systems in theory \cite{rashba:2003a,duckheim:2006a}
and experiment \cite{kato:2004a}, and can also be applied to lower-dimensional
systems (quantum wires and quantum dots) \cite{levitov:2003a,golovach:2006a},
with the advantage that single-qubit operations could then be performed
using fast all-electrical control. New experiments have now shown
that it may be possible in practice to perform single-spin operations
on single quantum dots using ESR, as depicted in Figure \ref{cap:Series-of-QDs}
\cite{koppens:2006a}.

As mentioned in Section \ref{sec:LossDiVincenzoOverview}, quantum-dot
spin readout can be performed using a spin filter. Experimentally,
spin filters have been reported in the open \cite{potok:2002a} and
Coulomb-blockade regimes \cite{folk:2003a}, and have even been used
to determine the longitudinal spin decay ($T_{1}$) time \cite{hanson:2003a,hanson:2004a}
using an $n$-shot readout scheme, which has been analyzed in detail
\cite{engel:2004a}. A single-shot readout has also been demonstrated
\cite{elzerman:2004a} and improved upon \cite{hanson:2005a}. Non-invasive
readout schemes using spin-to-charge conversion and quantum-point-contact
(QPC) measurements have been used on two-spin encoded qubits \cite{johnson:2005a,petta:2005a,petta:2005b,johnson:2005b}. 

To measure the transverse spin coherence time $T_{2}$, there have
been proposals to perform ESR and detect the resulting resonance in
stationary current \cite{engel:2001a}, changes in the resistivity
of a neighboring field-effect transistor (FET) \cite{martin:2003a},
optically \cite{gywat:2004a}, or from current noise \cite{schaefers:2005a}.
ESR in single quantum dots has not yet been observed, in part because
it is challenging to generate high-frequency magnetic fields with
sufficient power for single-spin manipulation without {}``heating''
electrons on the quantum dot or in the surrounding leads through the
associated electric field \cite{vanderwiel:2006a}. Recent experiments
that employ a double quantum dot in the spin-blockade regime may have
overcome this problem \cite{koppens:2006a} (see also the discussion
on spin blockade near the end of Section \ref{sec:Decoherence} below).

\section{\label{sec:DoubleQDs}DOUBLE QUANTUM DOTS}

Single qubits are the fundamental unit of quantum information in quantum
computing. However, universal quantum computation still requires both
single-qubit and two-qubit operations \cite{barenco:1995a}. In the
Loss-DiVincenzo proposal, two-qubit gates are performed with exchange-coupled
electron spins confined to two neighboring quantum dots (double dots).
Double dots are also important for encoded qubits \cite{levy:2002a},
in which qubits are encoded into a two-dimensional pseudospin-$1/2$
subspace of a four-dimensional two-electron spin system.

In this section we discuss characterization and manipulation techniques
that are commonly used to extract microscopic parameters of double
quantum dots. In Section \ref{sub:ChargeStabilityDiagram} we review
the charge stability diagram, and illustrate its connection to a commonly
used microscopic model Hamiltonian. In Section \ref{sub:MolecularStates}
we review work on the coherent coupling of double quantum dots, which
is required to generate a large exchange interaction for two-qubit
gating. In Section \ref{sub:TwoQubitGates} we discuss the use of
double quantum dots as two-qubit gates, and in Section \ref{sub:EncodedQubits}
we review some work on using double quantum dots to control single
{}``encoded'' qubits \cite{levy:2002a}, a topic which has now come
into vogue \cite{petta:2005a,taylor:2005c,burkard:2006a,hanson:2006a}.

\subsection{\label{sub:ChargeStabilityDiagram}The double-dot charge stability
diagram}

Just as transport through a single quantum dot and Coulomb blockade
phenomena give information about the orbital level spacing, charging
energy, and spin states of single quantum dots, similar studies can
be carried-out on double quantum dots. Whereas for single dots, transport
phenomena are typically understood in terms of one-dimensional plots
of conductance versus gate voltage, the primary tool used to understand
double quantum dots is the double-dot charge stability diagram. The
stability diagram is a two-dimensional plot of current or differential
conductance through the double dot or through a neighboring QPC, given
as a function of two independent back-gate voltages (one applied locally
to each dot). The plot differentiates regions where the double-dot
ground state has a charge configuration $(N_{1},N_{2})$, for various
$N_{1},N_{2}$, where $N_{1}$ is the number of charges on the left
dot and $N_{2}$ is the number of charges on the right. Transport
through double quantum dots and the relevant charge stability diagram
has been discussed thoroughly in \cite{vanderwiel:2003a}. In the
rest of this section, we review some features of the double-dot stability
diagram with an emphasis on the connection to a model Hamiltonian
that is commonly used in the literature \cite{klimeck:1994a,pals:1996a,golden:1996a,ziegler:2000a}. 

An isolated double quantum dot is described by the Hamiltonian\begin{equation}
H_{\mathrm{dd}}=H_{\mathrm{C}}+H_{\mathrm{T}}+H_{\mathrm{S}},\label{eq:HddDefinition}\end{equation}
where $H_{\mathrm{C}}$ gives the single-particle and inter-particle
charging energies as well as the orbital energy, $H_{\mathrm{T}}$
is the inter-dot tunneling term due to a finite overlap of dot-localized
single-particle wavefunctions, which ultimately gives rise to exchange,
and $H_{\mathrm{S}}$ contains explicitly spin-dependent terms, which
may include spin-orbit interaction, dipole-dipole interaction, and
the contact hyperfine interaction between the confined electron spins
and nuclear spins in the surrounding lattice. 

There are several approaches that can be taken to writing the various
components of the double-dot Hamiltonian $H_{\mathrm{dd}}$, corresponding
to several degrees of microscopic detail. In the simplest form, the
Hubbard model, details of the electron wavefunctions are neglected
and the Coulomb interaction is given only in terms of on-site and
nearest-neighbor terms. Since this description relies only on very
few parameters, it is the most commonly used in the literature on
transport phenomena through quantum dots. The shape of the confining
potential, quantum-dot localized wavefunctions, and form of the Coulomb
interaction may become important in certain circumstances, in which
case it is more appropriate to apply either the Heitler-London method
(which neglects doubly-occupied dot levels), or the Hund-Mulliken
method, which includes the effects of double-occupancy. These methods
predict, for instance, a variation of the interdot exchange interaction
through zero with increasing out-of-plane magnetic field \cite{burkard:1999a}.
Experimentally, it has been confirmed that the exchange coupling can
be tuned with an out-of-plane magnetic field in single vertical \cite{fujisawa:2002a}
and single lateral quantum dots \cite{zumbuhl:2004a}, which behave
effectively as double-dot structures. Here we ignore these effects
and focus on the simplest Hubbard model that reproduces much of the
double-dot physics that can be seen in transport phenomena. 

\begin{figure}
\begin{center}\includegraphics[%
  scale=0.5]{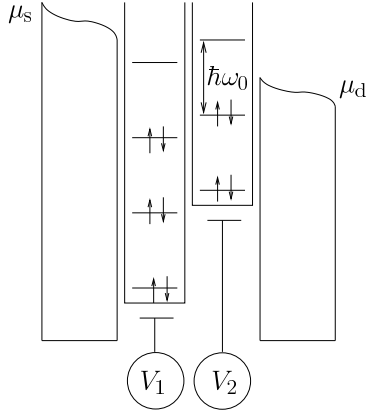}\end{center}

\caption{\label{cap:DoubleDotGroundState}Ground-state configuration for a
double quantum dot with large orbital and charging energies, and negligible
dot-lead and interdot coupling. $\mu_{\mathrm{s}(\mathrm{d})}$ is
the source (drain) chemical potential, $V_{1(2)}$ is the left (right)
local dot potential, which is related to applied gate potentials by
a linear transformation (see Equation (\ref{eq:V12Vg12Conversion}),
below), and both dots are assumed to have the same uniform level spacing
$\hbar\omega_{0}$.}
\end{figure}

\begin{figure}
\begin{center}\includegraphics[%
  scale=0.25]{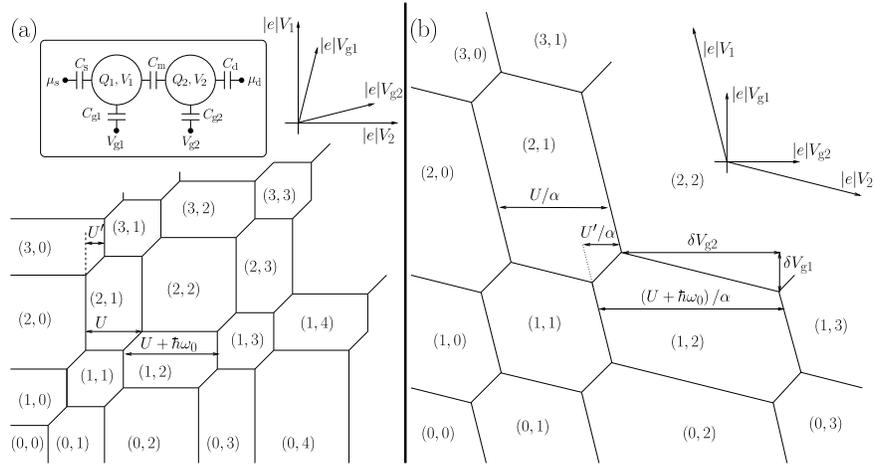}\end{center}

\caption{\label{cap:Stability-diagram-plotted}Stability diagram plotted in
terms of (a) local dot potentials $V_{1,2}$ and (b) applied gate
potentials $V_{g1,2}$, with on-site charging energies $U_{l}=U,\, l=1,2$,
nearest-neighbor charging energy $U^{\prime}$, and dot orbital level
spacing $\hbar\omega_{0}$ satisfying $U:\hbar\omega_{0}:U^{\prime}=3:2:1$.
In addition, for (b) we have assumed the voltage scaling factors are
the same for both dots, and are given by $\alpha_{1}=\alpha_{2}=\alpha=1/2$.
(a) inset: capacitive charging model for a double quantum dot, indicating
the source (drain) chemical potential $\mu_{\mathrm{s(d)}}$, the
charge on the left (right) dot $Q_{1(2)}$, the capacitances to source
(drain) $C_{\mathrm{s}(\mathrm{d})}$, the mutual capacitance $C_{\mathrm{m}}$,
and gate capacitances $C_{\mathrm{g}1,2}$. (b) Horizontal lines in
the $\left|e\right|V_{1(2)}$ plane become skewed with slope $\delta V_{\mathrm{g1}}/\delta V_{\mathrm{g}2}=-C_{1}C_{\mathrm{g}2}/C_{\mathrm{m}}C_{\mathrm{g}1}$
when plotted versus $\left|e\right|V_{\mathrm{g1(2)}}$.}
\end{figure}
We model the Coulomb interaction with simple on-site ($U_{1(2)}$
for the left (right) dot) and nearest-neighbor ($U^{\prime}$) repulsion.
The single-particle charging energy is given in terms of a local dot
potential $V_{1(2)}$. The charging Hamiltonian is then\begin{equation}
H_{\mathrm{C}}=\frac{1}{2}\sum_{l}U_{l}N_{l}\left(N_{l}-1\right)+U^{\prime}N_{1}N_{2}-\left|e\right|\sum_{l}V_{l}N_{l}+\sum_{kl}\epsilon_{lk}n_{lk},\label{eq:HcDefinition}\end{equation}
 where $N_{l}=\sum_{k}n_{lk}$ counts the total number of electrons
in dot $l$, with $n_{lk}=\sum_{\sigma}d_{lk\sigma}^{\dagger}d_{lk\sigma}$,
and here $d_{lk\sigma}$ annihilates an electron on dot $l$, in orbital
$k$, with spin $\sigma$. $\epsilon_{lk}$ is the energy of single-particle
orbital level $k$ in dot $l$, which gives rise to the typical orbital
level spacing $\epsilon_{lk+1}-\epsilon_{lk}\approx\hbar\omega_{0}$
(see Figure \ref{cap:DoubleDotGroundState}). 

Within the capacitive charging model described by the equivalent circuit
in the inset of Figure \ref{cap:Stability-diagram-plotted}(a), the
microscopic charging energies are related to capacitances by \cite{ziegler:2000a,vanderwiel:2003a} 

\begin{equation}
U_{l}=\frac{C_{1}C_{2}}{C_{1}C_{2}-C_{\mathrm{m}}^{2}}\frac{e^{2}}{C_{l}},\,\,\,\,\, U^{\prime}=\frac{2e^{2}C_{\mathrm{m}}}{C_{1}C_{2}-C_{\mathrm{m}}^{2}},\end{equation}
where $C_{1}=C_{\mathrm{s}}+C_{\mathrm{m}}+C_{\mathrm{g}1}$, $C_{2}=C_{\mathrm{d}}+C_{\mathrm{m}}+C_{\mathrm{g}2}$,
and all capacitances are defined in the inset of Figure \ref{cap:Stability-diagram-plotted}(a).
In experiments, the local quantum dot potentials $V_{1,2}$ are controlled
indirectly in terms of gate voltages $V_{\mathrm{g}1,2}$, which are
capacitively coupled to the dots through gate capacitances $C_{\mathrm{g}1,2}$.
For fixed quantum-dot charges $\left(Q_{1},Q_{2}\right)=-\left|e\right|\left(N_{1},N_{2}\right)=\mathrm{const.}$,
differences in the dot voltages $\Delta V_{1}$ and $\Delta V_{2}$
are related to differences in the gate voltages $\Delta V_{\mathrm{g}1}$
and $\Delta V_{\mathrm{g}2}$ through \cite{ziegler:2000a,vanderwiel:2003a}\begin{equation}
\left(\begin{array}{cc}
C_{1} & -C_{\mathrm{m}}\\
-C_{\mathrm{m}} & C_{2}\end{array}\right)\left(\begin{array}{c}
\Delta V_{1}\\
\Delta V_{2}\end{array}\right)=\left(\begin{array}{c}
C_{\mathrm{g}1}\Delta V_{\mathrm{g}1}\\
C_{\mathrm{g}2}\Delta V_{\mathrm{g}2}\end{array}\right).\label{eq:V12Vg12Conversion}\end{equation}
The double-dot stability diagram can then be given equivalently as
a two-dimensional plot with energy axes $\left|e\right|V_{1}, \left|e\right|V_{2}$,
or with axes $\left|e\right|V_{\mathrm{g}1}, \left|e\right|V_{\mathrm{g}2}$,
which are skewed and stretched with respect to the original axes according
to the transformation given in Equation (\ref{eq:V12Vg12Conversion}).
The end effect is that parallel horizontal (vertical) lines in the
$\left|e\right|V_{1(2)}$ plane separated by a distance $dV_{2(1)}$
transform to skewed parallel lines, separated by $dV_{\mathrm{g}2(1)}=dV_{2(1)}/\alpha_{2(1)}$
along the horizontal (vertical) of the new coordinate system, where
(see Figure \ref{cap:Stability-diagram-plotted}):\begin{equation}
\alpha_{l}=\frac{C_{\mathrm{g}l}}{C_{l}},\,\,\,\,\, l=1,2.\label{eq:ScaleFactors}\end{equation}
Additionally, horizontal lines in the $\left|e\right|V_{1(2)}$ plane
become skewed with a slope $\delta V_{\mathrm{g}1}/\delta V_{\mathrm{g}2}=-C_{\mathrm{m}}C_{\mathrm{g}2}/C_{2}C_{\mathrm{g}1}$
(see Figure \ref{cap:Stability-diagram-plotted}(b)), and vertical
lines are skewed with slope $\delta V_{\mathrm{g}1}/\delta V_{\mathrm{g}2}=-C_{1}C_{\mathrm{g}2}/C_{\mathrm{m}}C_{\mathrm{g}1}$. 

The Hamiltonian in Equation (\ref{eq:HcDefinition}) conserves the
number of electrons on each dot: $\commute{H_{\mathrm{C}}}{N_{l}}=0$,
so we label the ground state by the two dot occupation numbers, $(N_{1},N_{2})$,
and indicate where each configuration is the ground state in Figure
\ref{cap:Stability-diagram-plotted} for equivalent quantum dots that
satisfy $\alpha_{1}=\alpha_{2}=\alpha=1/2$, $U_{1}=U_{2}=U$, $\epsilon_{lk+1}-\epsilon_{lk}=\hbar\omega_{0}$
for all $k,l$, and $U:\hbar\omega_{0}:U^{\prime}=3:2:1$. The charge
stability diagram produces a {}``honeycomb'' of hexagons with dimensions
that are determined by three typical energy scales: (1) The on-site
replusion $U$, (2) the nearest-neighbor repulsion $U^{\prime}$,
and (3) the typical orbital energy $\hbar\omega_{0}$. Figure \ref{cap:Stability-diagram-plotted}
assumes a ground-state electron filling as shown in Figure \ref{cap:DoubleDotGroundState},
with constant orbital energy $\hbar\omega_{0}$. In this case, the
orbital energy appears in the dimensions of only every second honeycomb
cell of the stability diagram, along the horizontal or vertical direction,
since the spin-degenerate orbital states fill with two electrons at
a time according to the Pauli principle. This even-odd behavior may
not be visible in dots of high symmetry, where the orbital levels
are manifold degenerate. Alternatively, the absence of an even-odd
effect in low-symmetry single dots has previously been attributed
to the absence of spin degeneracy due to many-body effects \cite{stewart:1997a,fujisawa:2001a,vanderwiel:2003a}.

Each vertex of a honeycomb cell corresponds to a triple-point, where
three double-dot charge states are simultaneously degenerate. For
a double dot connected to source and drain leads at low temperature,
and in the absence of relaxation or photo-assisted tunneling processes,
it is only at these points where resonant sequential transport can
occur, through shuttling processes of the form $(0,0)\to(1,0)\to(0,1)\to(0,0)$.
This picture changes when a strong inter-dot tunnel coupling $H_{\mathrm{T}}$
is considered in addition.

\subsection{\label{sub:MolecularStates}Molecular states in double dots }

\begin{figure}
\begin{center}\includegraphics[%
  scale=0.5]{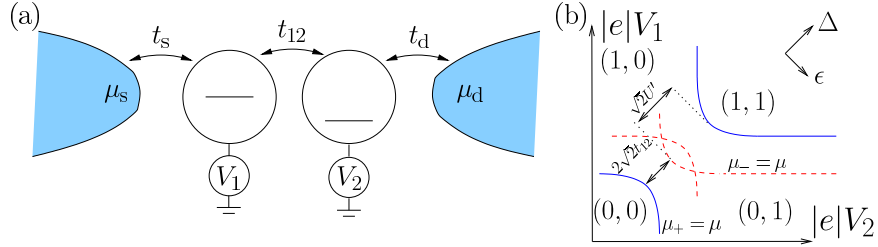}\end{center}

\caption{\label{cap:Tunnel-Coupled-double-dot} (a) A tunnel-coupled double
quantum dot, with tunneling amplitude $t_{12}$. The source and drain
leads, at chemical potentials $\mu_{\mathrm{s}}$ and $\mu_{\mathrm{d}}$,
are connected to the left and right dots through tunnel barriers with
tunneling amplitudes $t_{\mathrm{s}}$ and $t_{\mathrm{d}}$, respectively.
The left and right dots are set to local potentials $V_{1}$ and $V_{2}$.
(b) Modification of the stability diagram in the case of a significant
tunnel coupling $t_{12}$. To generate this figure we have chosen
the ratio of tunnel coupling to the mutual (nearest-neighbor) charging
energy to be $t_{12}/U^{\prime}\approx1/5$. At solid lines, transport
occurs via the double-dot ground state $\ket{E_{+}}$ and at dashed
lines, additional transport can occur through the first excited state
$\ket{E_{-}}$ (see Equations (\ref{eq:EpmStates}) and (\ref{eq:EpmMixingAngle})
below).}
\end{figure}

Molecule-like states have been observed and studied in detail in two-electron
single vertical \cite{fujisawa:2002a} and lateral quantum dots \cite{zumbuhl:2004a}
(the latter behave as an effective double-dot structure, showing good
agreement with theory \cite{golovach:2004b}). Evidence of molecular
states forming in double quantum dots due to a strong inter-dot tunnel-coupling
has also been found in a variety of systems \cite{schmidt:1997a,schedelbeck:1997a,blick:1998a,brodsky:2000a,bayer:2001a,ota:2005a,huettel:2005a,fasth:2005a,mason:2004a,biercuk:2005a,graeber:2006a}.
For example, molecular states have been observed in many-electron
gated quantum dots in linear transport \cite{blick:1998a} (solid
lines of Figure \ref{cap:Tunnel-Coupled-double-dot}(b)) and transport
through excited states \cite{huettel:2005a} (dashed lines in Figure
\ref{cap:Tunnel-Coupled-double-dot}(b)). In addition, molecular states
have been observed in vertical-lateral gated double quantum dots \cite{hatano:2005a},
gated dots formed in quantum wires \cite{fasth:2005a} and gated carbon-nanotube
double dots \cite{mason:2004a,biercuk:2005a,graeber:2006a}. A large
inter-dot tunnel coupling is essential for generating a large exchange
interaction $J$, and is therefore very important for the implementation
of fast two-qubit gates in the Loss-DiVincenzo proposal.

In this section, we analyze changes to the double-dot stability diagram
that occur due to the inter-dot tunneling term $H_{\mathrm{T}}$.
We focus on the relevant regime for quantum computing, where only
a single orbital state is available for occupation on each quantum
dot (the lower-left region of Figures \ref{cap:Stability-diagram-plotted}(a,b)).
In the subspace of these lowest dot orbital states, $H_{\mathrm{T}}$
is given by:\begin{equation}
H_{\mathrm{T}}=\sum_{\sigma}t_{12}d_{1\sigma}^{\dagger}d_{2\sigma}+\mathrm{H.c.},\label{eq:HTDefinition}\end{equation}
 where $t_{12}$ is the tunneling amplitude between the two dots,
and $d_{l\sigma}$, $l=1,2$, annihilates an electron in the lowest
single-particle orbital state localized on quantum dot $l$ with spin
$\sigma$.

When the double dot is occupied by only $N=0,1$ electrons and is
coupled weakly to leads, an explicit expression can be found for the
current passing through a sequentially-coupled double dot, as shown
in Figure \ref{cap:Tunnel-Coupled-double-dot}(a) \cite{ziegler:2000a,graeber:2006a}.
It is straightforward to diagonalize $H_{C}+H_{T}$ in the subspace
of $N=1$ electrons on the quantum dot. This gives the (spin-degenerate)
eigenenergies and corresponding eigenvectors:\begin{eqnarray}
E_{\pm}(\Delta,\epsilon) & = & -\frac{1}{\sqrt{2}}\left(\Delta\pm\sqrt{\epsilon^{2}+2t_{12}^{2}}\right),\label{eq:EpmEigenvalues}\\
\ket{E_{\pm}} & = & \cos\left(\frac{\theta_{\pm}}{2}\right)\ket{1,0}+\sin\left(\frac{\theta_{\pm}}{2}\right)\ket{0,1},\label{eq:EpmStates}\\
\tan\left(\frac{\theta_{\pm}}{2}\right) & = & \frac{\epsilon}{\sqrt{2}t_{12}}\pm\sqrt{1+\left(\frac{\epsilon}{\sqrt{2}t_{12}}\right)^{2}}.\label{eq:EpmMixingAngle}\end{eqnarray}
 Here, $E_{\pm}(\Delta,\epsilon)$ is written in terms of new energy
coordinates $\epsilon,\,\Delta$, which are related to the old (voltage)
coordinates through a rotation of the axes by $45^{\circ}$ (see also
Figure \ref{cap:Tunnel-Coupled-double-dot}(b)): \begin{equation}
\left(\begin{array}{c}
\Delta\\
\epsilon\end{array}\right)=\frac{1}{\sqrt{2}}\left(\begin{array}{cc}
1 & 1\\
-1 & 1\end{array}\right)\left(\begin{array}{c}
\left|e\right|V_{1}\\
\left|e\right|V_{2}\end{array}\right).\end{equation}
We then define double-dot chemical potentials:\begin{equation}
\mu_{\pm}(\Delta,\epsilon)=E_{\pm}(\Delta,\epsilon)-E_{0},\label{eq:mupmDefinition}\end{equation}
where $E_{0}=0$ is the energy of the $(0,0)$ charge configuration.
In the presence of a strong tunnel coupling, the eigenstates of the
double dot are no longer labeled separately by the quantum numbers
$N_{1}, N_{2}$. Instead, the sum $N=N_{1}+N_{2}$ is conserved. If
we add to $H_{\mathrm{dd}}$ the double-dot-lead coupling Hamiltonian
$H_{\mathrm{dd}-L}=\sum_{k\sigma}t_{\mathrm{s}}c_{\mathrm{s}k\sigma}^{\dagger}d_{1\sigma}+t_{\mathrm{d}}c_{\mathrm{d}k\sigma}^{\dagger}d_{2\sigma}+\mathrm{H.c.}$,
where $c_{\mathrm{s(d)}k\sigma}^{\dagger}$ creates an electron in
the source (drain), in orbital $k$ with spin $\sigma$, then $N$
can fluctuate between $1$ and $0$ if the double-dot and lead chemical
potentials are equal. We identify double-dot sequential tunneling
processes as those that change the total charge on the double dot
by one: $N\to N\pm1$ \cite{golovach:2004b}. One can evaluate golden-rule
rates for all sequential-tunneling processes, taking the dot-lead
coupling $H_{\mathrm{dd}-\mathrm{L}}$ as a perturbation to obtain
the stationary current from a standard Pauli master equation (the
Pauli master equation is valid for sufficiently high temperature,
$k_{\mathrm{B}}T>\Gamma_{\mathrm{s(d)}}$, so that off-diagonal elements
can be ignored in the double-dot density matrix). For weak dot-lead
coupling, at low temperature $k_{\mathrm{B}}T<\hbar\omega_{0}$, and
at zero bias ($\mu=\mu_{s}=\mu_{d}+\Delta\mu$, with $\Delta\mu\to0$),
transport occurs only through the $N=1$ ground state, with chemical
potential $\mu_{+}$. The differential conductance near the $N=0,1$
boundary is then given by\begin{equation}
\frac{dI}{d\left(\Delta\mu\right)}=\left|e\right|\Gamma\left(\frac{-2f^{\prime}(\mu_{+})}{1+f(\mu_{+})}\right),\,\,\,\,\,\Gamma=\frac{\sin^{2}\left(\theta_{+}\right)\Gamma_{\mathrm{s}}\Gamma_{\mathrm{d}}}{4\left(\cos^{2}\left(\frac{\theta_{+}}{2}\right)\Gamma_{\mathrm{s}}+\sin^{2}\left(\frac{\theta_{+}}{2}\right)\Gamma_{\mathrm{d}}\right)},\label{eq:DDDiffCond}\end{equation}
 where $f(E)=1/\left[1+\exp\left(\frac{E-\mu}{k_{B}T}\right)\right]$
is the Fermi function at chemical potential $\mu$ and temperature
$T$, $f^{\prime}(E)=df(E)/dE$, and $\Gamma_{\mathrm{s(d)}}=\frac{2\pi\nu}{\hbar}\left|t_{\mathrm{s(d)}}\right|^{2}$
is the tunneling rate to the source (drain) with a lead density of
states per spin $\nu$ at the Fermi energy. If spin degeneracy is
lifted, the quantity in brackets in Equation (\ref{eq:DDDiffCond})
is replaced by the familiar term $-f^{\prime}(\mu_{+})=1/\left[4k_{B}T\cosh^{2}\left(\frac{\mu_{+}-\mu}{2k_{B}T}\right)\right]$
\cite{beenakker:1991a}. The differential conductance (Equation (\ref{eq:DDDiffCond}))
reaches a maximum near the point where the double-dot chemical potential
matches the lead chemical potential, $\mu_{+}(\Delta,\epsilon)=\mu$,
which we indicate with a solid line in Figure \ref{cap:Tunnel-Coupled-double-dot}(b).
Transport through the excited state can occur where $\mu_{-}(\Delta,\epsilon)=\mu$,
and when the bias $\Delta\mu=\mu_{\mathrm{s}}-\mu_{\mathrm{d}}$ or
temperature $T$ are sufficiently large to generate a significant
population in the excited state $\ket{E_{-}}$. Dashed lines indicate
where $\mu_{-}(\Delta,\epsilon)=\mu$ in Figure \ref{cap:Tunnel-Coupled-double-dot}(b).

There are several qualitative changes to the double-dot stability
diagram that take place in the presence of strong tunnel coupling.
First, the number of electrons on each dot is not conserved individually.
Instead, the sum $N=N_{1}+N_{2}$ is conserved, which means that there
are no longer lines separating, for example, the (1,0) and (0,1) states
in Figure \ref{cap:Tunnel-Coupled-double-dot}(b). Second, sequential-tunneling
processes allow current to be transported through the double-dot along
the length of the {}``wings'' that define the boundaries between
$N$ and $N\pm1$-electron ground states. This is in contrast to the
case where $t_{12}$ is weak, in which resonant sequential transport
can only occur at triple points, where the shuttling processes of
the type $(0,0)\to(1,0)\to(0,1)\to(0,0)$ are allowed by energy conservation.

\subsection{\label{sub:TwoQubitGates}Double dots as two-qubit gates}

The $\sqrt{\mbox{\sc swap}}$ operation described in Section \ref{sec:LossDiVincenzoOverview}
requires significant control of the exchange coupling $J$. The value
of $J$ can be controlled by raising/lowering the inter-dot barrier,
thus changing the tunnel coupling $t_{12}$ \cite{loss:1998a}, or
with an out-of-plane magnetic field or weak in-plane electric field
\cite{burkard:1999a}. More recently, experiments have controlled
$J$ by varying the back-gate voltages on two neighboring quantum
dots through a large parameter regime, independently \cite{petta:2005a}.
Here we discuss this last method to control $J$, which has been analyzed
in several recent papers \cite{petta:2005b,coish:2005a,taylor:2006a,stopa:2006a}. 

\begin{figure}
\begin{center}\includegraphics{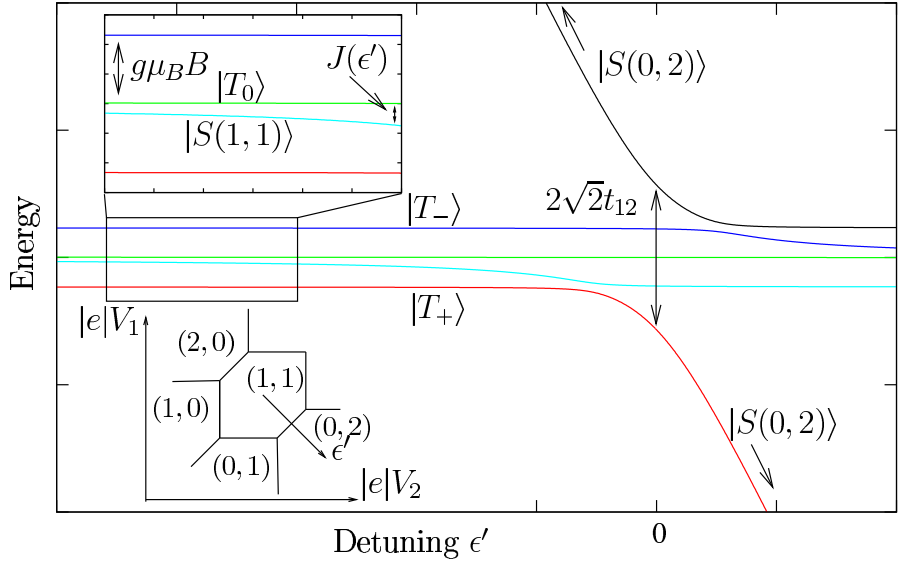}\end{center}

\caption{\label{cap:TwoElectronEnergySpectrum}Energy-level spectrum for two
electrons in a double quantum dot.}
\end{figure}
We consider a double quantum dot in the region of the charge stability
diagram indicated in the lower inset of Figure \ref{cap:TwoElectronEnergySpectrum}.
Specifically, we consider the regime of gate voltages where the double
dot contains $N=2$ electrons near the degeneracy point of the $(1,1)$
and $(0,2)$ charge states, and aim to diagonalize the Hamiltonian
$H_{\mathrm{C}}+H_{\mathrm{T}}$ in the basis of three spin triplets
and two relevant singlets:\begin{eqnarray}
\ket{S(0,2)} & = & d_{2\downarrow}^{\dagger}d_{2\uparrow}^{\dagger}\ket{\mathrm{vac.}},\label{eq:S02Definition}\\
\ket{S(1,1)} & = & \frac{1}{\sqrt{2}}\left(d_{2\downarrow}^{\dagger}d_{1\uparrow}^{\dagger}-d_{2\uparrow}^{\dagger}d_{1\downarrow}^{\dagger}\right)\ket{\mathrm{vac.}},\label{eq:S11Definition}\\
\ket{T_{0}} & = & \frac{1}{\sqrt{2}}\left(d_{2\downarrow}^{\dagger}d_{1\uparrow}^{\dagger}+d_{2\uparrow}^{\dagger}d_{1\downarrow}^{\dagger}\right)\ket{\mathrm{vac.}},\label{eq:T0Definition}\\
\ket{T_{+}} & = & d_{2\uparrow}^{\dagger}d_{1\uparrow}^{\dagger}\ket{\mathrm{vac.}},\label{eq:TplusDefinition}\\
\ket{T_{-}} & = & d_{2\downarrow}^{\dagger}d_{1\downarrow}^{\dagger}\ket{\mathrm{vac.}}.\label{eq:TminusDefinition}\end{eqnarray}
In the absence of additional spin-dependent terms, the triplets are
degenerate, with energy $E_{\mathrm{Triplet}}=E_{(1,1)}=-\sqrt{2}\Delta^{\prime}$,
whereas the two singlet states have energies and associated eigenvectors
\begin{eqnarray}
E_{\mathrm{Singlet}}^{\pm} & = & E_{\mathrm{Triplet}}-\frac{1}{\sqrt{2}}\left(\epsilon^{\prime}\pm\sqrt{\left(\epsilon^{\prime}\right)^{2}+4t_{12}^{2}}\right),\label{eq:Esinglet}\\
\ket{E_{\mathrm{Singlet}}^{\pm}} & = & \cos\left(\frac{\theta_{\pm}^{S}}{2}\right)\ket{S(1,1)}+\sin\left(\frac{\theta_{\pm}^{S}}{2}\right)\ket{S(0,2)},\label{eq:EsingletEigenstates}\\
\tan\left(\frac{\theta_{\pm}^{S}}{2}\right) & = & \frac{\epsilon^{\prime}}{2t_{12}}\pm\sqrt{1+\left(\frac{\epsilon^{\prime}}{2t_{12}}\right)^{2}}.\label{eq:EsingletMixingAngle}\end{eqnarray}
Here, $\Delta^{\prime}$ and $\epsilon^{\prime}$ are related to the
previous coordinates $\left(\Delta,\,\epsilon\right)$ through a simple
translation of the origin:\begin{equation}
\left(\begin{array}{c}
\Delta^{\prime}\\
\epsilon^{\prime}\end{array}\right)=\left(\begin{array}{c}
\Delta\\
\epsilon\end{array}\right)+\frac{1}{\sqrt{2}}\left(\begin{array}{c}
-U^{\prime}\\
U^{\prime}-U\end{array}\right).\end{equation}
 This gives rise to the Heisenberg exchange for large negative $\epsilon^{\prime}$
(from Equation (\ref{eq:Esinglet})):\begin{equation}
J(\epsilon^{\prime})=E_{\mathrm{Triplet}}-E_{\mathrm{Singlet}}^{+}\approx\frac{\sqrt{2}t_{12}^{2}}{\left|\epsilon^{\prime}\right|},\,\,\,\,\epsilon^{\prime}<0,\,\,\,\,\left|\epsilon^{\prime}\right|\gg2t_{12}.\label{eq:JLargeEpsilon}\end{equation}
By pulsing $\epsilon^{\prime}=\epsilon^{\prime}(t)$, the exchange
$J(\epsilon^{\prime}(t))$ can be pulsed on and off again in order
to implement the $\sqrt{\mbox{{\sc SWAP}}}$ operation, as described
in Section \ref{sec:LossDiVincenzoOverview} (see the inset of Figure
\ref{cap:TwoElectronEnergySpectrum}). This operation has now been
achieved experimentally with a gating time on the order of $180\,\mathrm{ps}$
\cite{petta:2005a}, in good agreement with the predictions in \cite{burkard:1999a}
for an achievable switching time.

\subsection{\label{sub:EncodedQubits}Initialization of two-spin encoded qubits}

Fluctuations in a nuclear spin environment can lead to rapid decoherence
of single-electron spin states due to the contact hyperfine interaction
(see Section \ref{sec:Decoherence}, below). The effects of these
fluctuations can be reduced, in part, by considering a qubit encoded
in two-electron singlet $\ket{0}=\ket{S(1,1)}$ and triplet states
$\ket{1}=\ket{T_{0}}$, as defined in equations (\ref{eq:S11Definition}),
(\ref{eq:T0Definition}). With this encoding scheme, the qubit energy
splitting would be provided through the exchange coupling (Equation
(\ref{eq:JLargeEpsilon})), and single-qubit rotations could be performed
using an inhomogeneous magnetic field \cite{levy:2002a}. Two-qubit
operations in this scheme could be performed, for example, using capacitive
coupling due to the relative charge distributions of the triplet and
singlet states in neighboring double-dots \cite{taylor:2005c}, although
the difference in these charge distributions can lead to additional
dephasing due to fluctuations in the electrical environment \cite{coish:2005a}
(see also Section \ref{sec:Decoherence}, below). An alternative scheme
to couple such encoded qubits over long distances with optical cavity
modes has also been proposed \cite{burkard:2006a}. One additional
advantage of the two-spin encoded qubit scheme is that adiabatic tuning
of the gate voltages can be used to initialize and readout information
stored in the singlet-triplet basis \cite{johnson:2005a,petta:2005a}.
We discuss this initialization scheme in the rest of this section.

We consider the singlet ground-state $\ket{E_{\mathrm{Singlet}}^{+}}$,
given by Equations (\ref{eq:EsingletEigenstates}) and (\ref{eq:EsingletMixingAngle}).
For large positive detuning, $\left|\epsilon^{\prime}\right|\gg t_{12},\epsilon^{\prime}>0$,
the mixing angle in Equation (\ref{eq:EsingletMixingAngle}) is $\theta_{+}^{S}\approx\pi$
and the singlet ground state is approximately given by $\ket{S(0,2)}$.
For large negative detuning $\left|\epsilon^{\prime}\right|\gg t_{12},\epsilon^{\prime}<0$,
we find $\theta_{+}^{S}\approx0$ and the lowest-energy singlet is
instead given by $\ket{S(1,1)}$ (see Figure \ref{cap:TwoElectronEnergySpectrum}).
If the two-electron system is allowed to relax to its ground state
$\ket{S(0,2)}$ at large positive detuning $\epsilon^{\prime}$ and
the detuning is then varied adiabatically slowly to large negative
values, the encoded qubit can be initialized to the state $\ket{0}=\ket{S(1,1)}$
(see Figure \ref{cap:TwoElectronEnergySpectrum} and insets). It is
a straightforward exercise to estimate the error in such an operation
for a two-dimensional Hamiltonian.

\begin{figure}
\begin{center}\includegraphics[%
  scale=0.7]{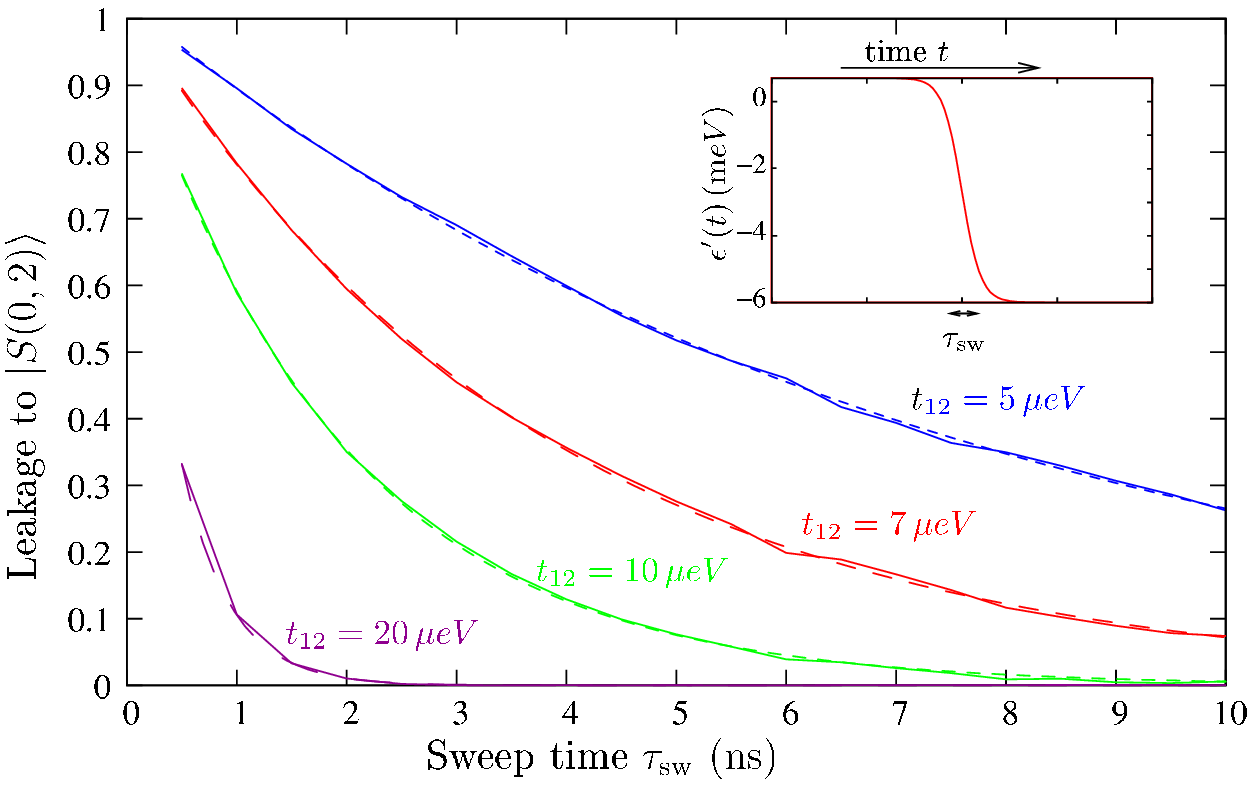}\end{center}

\caption{\label{cap:LandauZenerTunnelingProbability}Leakage (the occupation
probability of the state $\ket{S(0,2)}$ at the end of the sweep)
due to nonadiabatic transitions after sweeping from $\epsilon^{\prime}=0.7\,\mathrm{m}eV$
to $\epsilon^{\prime}=-6\,\mathrm{m}eV$. Leakage is given as a function
of the characeristic sweep time $\tau_{\mathrm{sw}}$, where $\epsilon^{\prime}(t)=\epsilon_{0}-\frac{\Delta\epsilon}{2}\tanh\left(2t/\tau_{sw}\right)$,
with $\epsilon_{0}=-2.65\,\mathrm{m}eV$ and $\Delta\epsilon=6.7\,\mathrm{m}eV$.
We show results for $t_{12}=5,7,10,20\,\mu eV$. Solid lines show
the results of numerical integration of the Schr\"odinger equation
and dashed lines give exponential fits, which decay with the time
constants $\tau_{\mathrm{fit}}$, given in Table \ref{cap:time-constant-table}. }
\end{figure}
For a linear ramp of $\epsilon^{\prime}$ over an infinite interval
(i.e., $\epsilon^{\prime}=\left(\Delta\epsilon/\tau_{\mathrm{sw}}\right)t,\, t=-\infty\ldots\infty$
with characteristic switching time $\tau_{\mathrm{sw}}$ to sweep
over an interval $\Delta\epsilon$), the result of Zener, for the
non-adiabatic Landau-Zener transition probability is\cite{zener:1932a}\begin{equation}
P=\exp\left(-\frac{\tau_{\mathrm{sw}}}{\tau_{LZ}}\right),\,\,\,\,\,\tau_{LZ}=\frac{\hbar\sqrt{2}\Delta\epsilon}{4\pi t_{12}^{2}}.\label{eq:ZenerTunnelingProbability}\end{equation}
Here, the Landau-Zener tunneling probability is controlled in terms
of two time scales: the switching time $\tau_{\mathrm{sw}}$ for a
typical range of $\Delta\epsilon$ and the Landau-Zener time constant
$\tau_{L\mathrm{Z}}$, which has a strong dependence ($\propto1/t_{12}^{2}$)
on the inter-dot tunnel coupling. For a realistic voltage pulse, $\epsilon^{\prime}$
is swept over a finite interval, and the pulse shape, in general,
will not be linear for the entire sweep. Performing an analysis similar
to that used for single-spin gates \cite{schliemann:2001a,requist:2005a}
for this case, one can perform a numerical integration of the time-dependent
Schr\"odinger equation in the subspace formed by the two singlets
for an arbitrary pulse shape. We have done this for a pulse of the
form $\epsilon^{\prime}(t)=\epsilon_{0}-\frac{\Delta\epsilon}{2}\tanh\left(2t/\tau_{\mathrm{sw}}\right),\, t=-5\tau_{\mathrm{sw}}\ldots5\tau_{\mathrm{sw}}$%
\footnote{Note that this type of pulse will generally lead to a smaller value
of $P$ for a given set of parameters since here, $d\epsilon^{\prime}/dt\leq\Delta\epsilon/\tau_{\mathrm{sw}}$,
whereas for the linear pulse $d\epsilon^{\prime}/dt=\Delta\epsilon/\tau_{\mathrm{sw}}$
for the entire sweep.%
}, where we find an approximately exponential dependence of $P$ on
the switching time $\tau_{\mathrm{sw}}$ (see Figure \ref{cap:LandauZenerTunnelingProbability}).
Fitting to this exponential dependence, we find a time constant $\tau_{\mathrm{fit}}$
analogous to the Landau-Zener time $\tau_{L\mathrm{Z}}$. The time
constants $\tau_{L\mathrm{Z}}$ (from Equation (\ref{eq:ZenerTunnelingProbability}))
and $\tau_{\mathrm{fit}}$ from the numerical data in Figure \ref{cap:LandauZenerTunnelingProbability}
are compared in Table \ref{cap:time-constant-table} for various values
of the inter-dot tunnel coupling $t_{12}$. The results of Figure
\ref{cap:LandauZenerTunnelingProbability} and Table \ref{cap:time-constant-table}
suggest that (for this set of parameters) adiabatic switching for
initialization or readout on a time scale of $\tau_{\mathrm{sw}}\lesssim1\,\mathrm{ns}$
can only be performed without significant error if the tunnel coupling
$t_{12}$ is made larger than $t_{12}>20\,\mu eV$. It is important
to note that this analysis ignores additional effects due to magnetic
field inhomogeneities, spin-orbit coupling, or the hyperfine interaction,
all of which can lead to additional singlet-triplet anticrossings
(see Section \ref{sub:Hyperfine-interaction}, below) and hence, to
additional initialization or readout errors. 

\begin{table}
\begin{center}\begin{tabular}{|c|c|c|}
\hline 
$t_{12}$~$(\mu eV)$&
$\tau_{LZ}\,\,(\mathrm{ns})$&
$\tau_{\mathrm{fit}}\,\,(\mathrm{ns})$\tabularnewline
\hline
\hline 
5&
20.&
7.4\tabularnewline
\hline 
7&
10.&
3.8\tabularnewline
\hline 
10&
4.9&
1.9\tabularnewline
\hline 
20&
1.2&
0.43\tabularnewline
\hline
\end{tabular}\end{center}

\caption{\label{cap:time-constant-table}Landau-Zener time constant $\tau_{LZ}$
for a linear ramp of $\epsilon^{\prime}$ and the time constant $\tau_{\mathrm{fit}}$
for fits to numerically evaluated data at various values of the tunnel
coupling $t_{12}$.}
\end{table}

\section{\label{sec:Decoherence}DECOHERENCE}

Decoherence is the process by which information stored in a quantum
bit is lost. There are two time scales used to describe decoherence
processes for a spin that decays exponentially in the presence of
an applied magnetic field. $T_{1}$ is the longitudinal spin decay
time, or spin-flip time, which describes the time scale for random
spin flips: $\ket{\uparrow}\to\ket{\downarrow}$. $T_{2}$, the transverse
spin decay time, describes the decay of a superposition state $a\ket{\uparrow}+b\ket{\downarrow}$.
Both of these time scales are important for quantum computing, since
both effects lead to qubit errors. 

An experiment performed on an ensemble of systems with different environments
can lead to additional decoherence, beyond that described by the {}``intrinsic''
$T_{2}$ time \cite{slichter:1980a}. For such an experiment, the
ensemble-averaged transverse spin decay time is therefore often denoted
$T_{2}^{*}$ to distinguish it from the single-spin decay time. Other
symbols such as $\tau_{c}$ (the correlation time) and $T_{\mathrm{M}}$
(the magnetization envelope decay time) are often used to distinguish
decay that is non-exponential. 

For a quantum-dot-confined electron spin state to decay, it is necessary
for the spin to couple in some way to fluctuations in the environment.
There are two important sources of this coupling for electron spins
in quantum dots. First, the spin-orbit interaction couples electron
spin states to their orbital states, and therefore makes spins indirectly
sensitive to fluctuations in the electric environment. Second, the
Fermi contact hyperfine interaction between electrons and surrounding
nuclear spins in the host material can lead to rapid decay if fluctuations
in the nuclear spin environment are not properly controlled. In the
rest of this section we discuss recent progress in understanding decoherence
due to these two coupling mechanisms.

\subsection{\label{sub:Spin-orbit-interaction}Spin-orbit interaction}

For a 2DEG formed in GaAs, the spin-orbit interaction is given in
terms of two terms:\begin{equation}
H_{\mathrm{SO}}=\alpha\left(p_{x}\sigma_{y}-p_{y}\sigma_{x}\right)+\beta\left(p_{y}\sigma_{y}-p_{x}\sigma_{x}\right)+O\left(\left|\mathbf{p}\right|^{3}\right),\label{eq:HsoDefinition}\end{equation}
 where $\sigma_{x,y}$ are Pauli matrices and $\mathbf{p}=(p_{x},p_{y})$
is the momentum operator in the plane of the 2DEG. The first term,
proportional to $\alpha$, is the Rashba (or structure-inversion-asymmetry)
spin-orbit coupling term. The Rashba term is due to asymmetry in the
confining potential and can therefore be tuned to some degree with
applied gates. The second term, proportional to $\beta$, is the Dresselhaus
(bulk-inversion-asymmetry) term, and is due to the fact that GaAs,
which has a zincblende lattice, has no center of inversion symmetry.
Corrections to this spin-orbit Hamiltonian of order $\left|\mathbf{p}\right|^{3}$
are smaller than the linear-momentum terms in quantum dots by the
ratio of $z$-confinement length to the quantum-dot Bohr radius, and
are negligible in the two-dimensional limit \cite{cerletti:2005a}. 

$H_{\mathrm{SO}}$ obeys time-reversal symmetry. Thus, in the absence
of a magnetic field, the ground state of a single electron confined
to a quantum dot is twofold degenerate due to Kramer's theorem, and
$H_{\mathrm{SO}}$ alone can not cause decoherence. The character
of the ground-state doublet does change, however, due to the presence
of $H_{\mathrm{SO}}$, mixing orbital and spin states. Thus, any fluctuations
that couple to the orbital degree of freedom can cause decoherence
in combination with spin-orbit coupling. These fluctuations can come
from lattice phonons, surrounding gates, electron-hole pair excitations,
etc. \cite{golovach:2004a}. The longitudinal-spin relaxation rate
$1/T_{1}$ due to spin-orbit coupling and lattice phonons has been
calculated, and shows a strong suppression for confined electrons
(with large level spacing $\hbar\omega_{0}$) in weak magnetic fields:
$1/T_{1}\propto B^{5}/\left(\hbar\omega_{0}\right)^{4}$ \cite{khaetskii:2000a,khaetskii:2001a}.
This calculation has been extended to a larger range of magnetic fields,
showing that the $B$-field dependence of $1/T_{1}$ saturates and
is then suppressed when the phonon wavelength is comparable to the
dot size \cite{golovach:2004a}. Further, this calculation has also
been extended to include the transverse spin decay time due to spin-orbit
interaction alone, showing that dephasing is limited by relaxation,
or $T_{2}=2T_{1}$ to leading order in the spin-orbit coupling, independent
of the particular source of fluctuations \cite{golovach:2004a}. Additionally,
$1/T_{1}$ has been shown to have a strong dependence on the magnetic
field direction, relative to the crystal axes \cite{falko:2005a},
shows a strong enhancement near avoided level crossings, which may
allow independent measurements of the Rashba and Dresselhaus coupling
constants \cite{bulaev:2005a}, and plays a role in phonon-assisted
cotunneling current through quantum dots \cite{lehmann:2006a}. The
relaxation rate of quantum-dot-confined hole spins due to spin-orbit
coupling and phonons has also been investigated. In some cases, recent
work has shown that the hole spin relaxation time may even exceed
the relaxation time of electron spins \cite{bulaev:2005b}. In addition
to lattice phonons, electric field fluctuations can result from the
noise in a QPC readout device, which results in spin decoherence when
considered in combination with spin-orbit coupling. This mechanism
shows a strong dependence of the decoherence rate $\sim1/r^{6}$ on
the dot-QPC separation $r$ \cite{borhani:2005a}, and can therefore
be controlled with careful positioning of the readout device. 

Measurements of relaxation times for single electron spins have been
performed in gated lateral quantum dots \cite{hanson:2003a,elzerman:2004a},
giving a $T_{1}$ time in good agreement with the theory of ref. \cite{golovach:2004a}
and in self-assembled quantum dots \cite{kroutvar:2004a}, which confirmed
the expected magnetic field dependence: $1/T_{1}\propto B^{5}$ \cite{khaetskii:2001a}.
Additionally, singlet-triplet decay has been measured in single vertical
\cite{fujisawa:2002a}, and lateral \cite{hanson:2005a} dots, as
well as lateral double dots \cite{petta:2005b,johnson:2005a}. 

There is a general consensus that spin relaxation for quantum-dot-confined
electrons proceeds through the spin-orbit interaction and phonon emission
at high magnetic fields. However, in weak magnetic fields, and for
the transverse spin decay time $T_{2}$, there are stronger effects
in GaAs. These effects are due to the contact hyperfine interaction
between confined electron spins and nuclear spins in the surrounding
lattice.

\subsection{\label{sub:Hyperfine-interaction}Hyperfine interaction}

For a collection of electrons in the presence of nuclear spins, the
Fermi contact hyperfine interaction reads

\begin{equation}
H_{\mathrm{hf}}=Av\sum_{k}\mathbf{I}_{k}\cdot\mathbf{S}(\mathbf{r}_{k}),\,\,\,\,\mathbf{S}(\mathbf{r}_{k})=\frac{1}{2}\sum_{\sigma,\sigma^{\prime}=\uparrow,\downarrow}\psi_{\sigma}^{\dagger}(\mathbf{r}_{k})\pmb{\sigma}_{\sigma\sigma^{\prime}}\psi_{\sigma^{\prime}}(\mathbf{r}_{k}),\label{eq:HhfGeneralDefinition}\end{equation}
where $A$ is the hyperfine coupling strength, $v$ is the volume
of a crystal unit cell containing one nuclear spin, $\mathbf{I}_{k}$
is the spin operator for the nuclear spin at site $k$, $\mathbf{S}(\mathbf{r}_{k})$
is the electron spin density at the nuclear site, given in terms of
field operators $\psi_{\sigma}(\mathbf{r})$ that satisfy the anticommutation
relations $\anticommute{\psi_{\sigma}(\mathbf{r})}{\psi_{\sigma^{\prime}}^{\dagger}(\mathbf{r}^{\prime})}=\delta(\mathbf{r}-\mathbf{r}^{\prime})\delta_{\sigma,\sigma^{\prime}},\,\,\anticommute{\psi_{\sigma}(\mathbf{r})}{\psi_{\sigma^{\prime}}(\mathbf{r}^{\prime})}=0,$
and we have denoted matrix elements by $\pmb{\sigma}_{\sigma\sigma^{\prime}}=\bra{\sigma}\pmb{\sigma}\ket{\sigma^{\prime}}$,
where $\pmb{\sigma}=(\sigma_{x},\sigma_{y},\sigma_{z})$ is the vector
of Pauli matrices. The significance of the general form given in Equation
(\ref{eq:HhfGeneralDefinition}) is that there can be an interplay
of orbital and spin degrees of freedom due to the contact hyperfine
interaction. When the orbital level spacing is not too large, this
interplay can be the limiting cause of electron-spin relaxation \cite{erlingsson:2001a,erlingsson:2002a}
in weak magnetic fields, where the spin-orbit interaction is less
effective, and leads to enhanced nuclear spin relaxation in the vicinity
of sequential-tunneling peaks for a quantum dot connected to leads,
where $\mathbf{S}(\mathbf{r})$ fluctuates significantly \cite{lyanda-geller:2002a,huttel:2004a}. 

The orbital level spacing in lateral quantum dots is usually much
larger than the typical energy scale of $H_{\mathrm{hf}}$. In this
case, it is possible to solve for the orbital envelope wavefunction
$\Psi_{0}(\mathbf{r})$ in the absence of the hyperfine interaction,
and write an effective hyperfine Hamiltonian for a single electron
confined to the quantum-dot orbital ground state:\begin{equation}
H_{\mathrm{hf},0}\approx\mathbf{h}_{0}\cdot\mathbf{S}_{0},\,\,\,\,\,\mathbf{h}_{0}=Av\sum\left|\Psi_{0}(\mathbf{r}_{k})\right|^{2}\mathbf{I}_{k},\label{eq:Hhf0Definition}\end{equation}
 where here $\mathbf{S}_{0}$ is the spin-$1/2$ operator for a single
electron in the quantum-dot orbital ground state. The primary material
used to make lateral quantum dots is GaAs. All natural isotopes of
Ga and As carry nuclear spin $I=3/2$. Each isotope has a distinct
hyperfine coupling constant, but the average coupling constant, weighted
by the relative abundance of each isotope in GaAs gives $A\approx90\,\mu eV$
\cite{paget:1977a}.

Dynamics under $H_{\mathrm{hf},0}$ have now been studied extensively
under many various approximations and in many parameter regimes. Here
we give a brief account of some part of this study. For an extensive
overview, see reviews in \cite{schliemann:2003a,cerletti:2005a}.
The first analysis of the influence of Equation (\ref{eq:Hhf0Definition})
on quantum-dot electron spin dynamics showed that the long-time longitudinal
spin-flip probability, $P_{\uparrow\downarrow}\approx1/p^{2}N$ \cite{burkard:1999a}
was suppressed in the limit of large nuclear spin polarization $p$
and number of nuclear spins in the dot, $N$. Subsequently, an exact
solution for the case of a fully-polarized nuclear spin system ($p=1$)
has shown that both the longitudinal and transverse components of
the electron spin decay by a fraction $\sim1/N$ according to a long-time
power law $\sim1/t^{3/2}$ on a time scale of $\tau\sim\hbar N/A$
\cite{khaetskii:2002a} ($\hbar N/A\sim1\,\mu\mathrm{s}$ for a GaAs
dot containing $N\simeq10^{5}$ nuclei). This exact solution for $p=1$,
which shows a non-exponential decay, demonstrates that the electron
spin decay is manifestly non-Markovian since the time scale for motion
in the nuclear-spin bath is much longer than the decay time scale
of the electron spin. For unpolarized systems, the ensemble averaged
mean-field dynamics show a transverse spin decay on a time scale $\tau\sim\hbar\sqrt{N}/A\sim5\,\mathrm{ns}$
\cite{khaetskii:2002a,merkulov:2002a}. The exact solution has been
extended to the case of nonzero polarization $p\ne1$ using a generalized
master equation, valid in the limit of large magnetic field or polarization
$p\gg1/\sqrt{N}$ \cite{coish:2004a}. This work has shown that, while
the longitudinal spin decay is bounded by $\sim1/p^{2}N$, due to
the quantum nature of the nuclear field, the transverse components
of spin will decay to zero in a time $t_{c}\approx5\,\mathrm{ns}/\sqrt{1-p^{2}}$
(without ensemble averaging and without making a mean-field ansatz),
unless an electron spin echo sequence is performed or the nuclei are
prepared in an eigenstate of the operator $h_{0}^{z}$ through measurement
\cite{coish:2004a}. There are several recent suggestions for methods
that could be used to measure the operator $h_{0}^{z}$ \cite{giedke:2005a,klauser:2005a,stepanenko:2005a}
in order to extend electron spin decoherence. Once the nuclear spin
system is forced into an eigenstate of $h_{0}^{z}$, the lowest-order
corrections for large magnetic field still show incomplete decay for
the transverse spin \cite{coish:2004a}, suggesting that dynamics
induced by the nuclear dipolar interaction may limit spin coherence
in this regime \cite{desousa:2003b}, although higher-order corrections
have been reported to lead to complete decay \cite{deng:2005a}, even
when the nuclear spin system is static. There have been several efforts
to understand the hyperfine decoherence problem numerically \cite{schliemann:2002a,shenvi:2005a},
and other studies have investigated electron spin-echo envelope decoherence
under the hyperfine interaction alone \cite{coish:2004a,shenvi:2005a,shenvi:2005b}
or the combined influence of hyperfine and nuclear dipolar interactions
\cite{desousa:2003b,desousa:2005a,witzel:2005a,yao:2005a,yao:2006a}.
Other approaches to understanding the hyperfine decoherence problem
include semiclassical theories that replace the quantum nuclear field
by a classical dynamical vector\cite{erlingsson:2004a,yuzbashyan:2004a}
or a classical distribution function \cite{alhassanieh:2005a}.

Experiments on electron spin decoherence in single quantum dots \cite{bracker:2005a,dutt:2005a}
and double quantum dots \cite{petta:2005a,koppens:2005a} have now
confirmed that the ensemble-averaged electron spin dephasing time
is indeed given by $\tau\sim\hbar\sqrt{N}/A\sim10\,\mathrm{ns}$.

For two electron spins confined to a double quantum dot, the hyperfine
Hamiltonian (Equation (\ref{eq:HhfGeneralDefinition})) can be cast
in the form \cite{coish:2005a}

\begin{eqnarray}
H_{\mathrm{hf},\mathrm{dd}} & = & \epsilon_{z}S_{l}^{z}+\sum_{l}\mathbf{h}_{l}\cdot\mathbf{S}_{l};\,\,\,\mathbf{S}_{l}=\frac{1}{2}\sum_{\sigma,\sigma^{\prime}=\uparrow,\downarrow}d_{l\sigma}^{\dagger}\pmb{\sigma}_{\sigma\sigma^{\prime}}d_{l\sigma^{\prime}},\\
 & = & \epsilon_{z}S_{l}^{z}+\mathbf{S}\cdot\mathbf{h}+\delta\mathbf{S}\cdot\delta\mathbf{h},\label{eq:HhfddDefinition}\end{eqnarray}
 where $\epsilon_{z}=g\mu_{\mathrm{B}}B$ is the Zeeman splitting,
$d_{1(2)\sigma}$ annihilates an electron in the single-particle orbital
state with envelope wavefunction $\Psi_{1(2)}(\mathbf{r})$ and spin
$\sigma$, we define $\mathbf{h}=\left(\mathbf{h}_{1}+\mathbf{h}_{2}\right)/2$,
$\delta\mathbf{h}=\left(\mathbf{h}_{1}+\mathbf{h}_{2}\right)/2$,
where the quantum nuclear field operators are $\mathbf{h}_{1(2)}=Av\sum_{k}\left|\Psi_{1(2)}(\mathbf{r}_{k})\right|^{2}\mathbf{I}_{k}$,
the sum of electron spins is $\mathbf{S}=\mathbf{S}_{1}+\mathbf{S}_{2}$
and the difference is $\delta\mathbf{S}=\mathbf{S_{1}-\mathbf{S}_{2}}$.
While the sum $\mathbf{S}$ conserves the total squared electron spin,
and can only couple states of different $z$-projection (e.g. $\ket{T_{0}}$
to $\ket{T_{\pm}}$), the difference $\delta\mathbf{S}$ does not
preserve the total spin, and therefore couples singlet to triplet
(e.g. $\ket{S(1,1)}$ to $\ket{T_{0}}$ and $\ket{T_{\pm}}$) and
will therefore lead to anticrossings in the energy level spectrum,
where $\ket{S(1,1)}$ and $\ket{T_{\pm}}$ or $\ket{T_{0}}$ cross.
Adding Equation (\ref{eq:HhfddDefinition}) to the previous double-dot
hamiltonian, $H_{\mathrm{dd}}=H_{\mathrm{C}}+H_{\mathrm{T}}+H_{\mathrm{hf},\mathrm{dd}}$,
and making a mean-field ansatz for the nuclear field operators, i.e.,
replacing operators by their expectation values: $\mathbf{h}\to\left\langle \mathbf{h}\right\rangle $,%
\footnote{In general, great care should be taken in making such a replacement.
See the discussion, for example, in \cite{coish:2005a}.%
} leads to the energy level spectrum shown in Figure \ref{cap:TwoElectronEnergySpectrum}.
In the limit of large Zeeman splitting $\epsilon_{z}$ and large negative
detuning $\epsilon^{\prime}$, an effective two-level Hamiltonian
can be derived in the subspace of lowest-energy singlet and $S^{z}=0$
triplet ($\ket{S},\ket{T_{0}}$) \cite{coish:2005a}:\begin{equation}
H_{\mathrm{dd},\mathrm{eff}}=\frac{J}{2}\mathbf{S}\cdot\mathbf{S}+\delta h^{z}\delta S^{z}+O\left(\frac{1}{\epsilon_{z}}\right).\label{eq:Heff}\end{equation}
 An exact solution can be found for pseudospin dynamics in the two-dimensional
subspace of $\ket{S}$ and $\ket{T_{0}}$ under the action of $H_{\mathrm{dd,eff}}$.
This solution shows that a singlet-triplet correlator undergoes an
interesting power-law decay in a characteristic time scale that can
be extended by increasing $J$ \cite{coish:2005a}, and has been verified
in experiment \cite{laird:2005a}. As is true for the transverse components
of a single electron spin, the singlet-triplet correlator shows a
rapid decay if the nuclear spin environment is not in an eigenstate
of the relevant nuclear field operator (in this case, $\delta h^{z}$).
The decay time can be significantly extended by narrowing the distribution
in $\delta h^{z}$ eigenstates through measurement \cite{klauser:2005a}
or by performing a spin-echo sequence \cite{petta:2005a}. Remaining
sources of dephasing include the corrections to $H_{\mathrm{dd},\mathrm{eff}}$
(of order $1/\epsilon_{z}$, which can not be removed easily) and
fluctuations in the electrostatic environment, although the effect
of these fluctuations can be removed to leading order at zero-derivative
points for the exchange interaction \cite{coish:2005a}, where:\begin{equation}
\frac{dJ(\epsilon)}{d\epsilon}=0.\end{equation}
 Recent calculations suggest that these zero-derivative points should
be achievable with appropriate control of the confinement potential
or magnetic field \cite{hu:2006a,stopa:2006a}.

Since the hyperfine interaction does not preserve the total spin quantum
number of electrons, this interaction plays a very important role
in studies on spin-dependent transport. In particular, spin blockade
\cite{weinmann:1995a,weinmann:2003a} occurs in double quantum dots
\cite{ono:2002a} when tunneling is allowed betwen spin-singlets $\ket{S(1,1)}\to\ket{S(0,2)}$,
but not between spin triplets $\ket{T(1,1)}\nrightarrow\ket{T(0,2)}$,
because of a large energy cost due to orbital level spacing and the
Pauli principle. This blockade allows for the extraction of features
at energy scales much less than temperature, making it an ideal parameter
regime in which to perform spectroscopy on double dots \cite{pioro-ladriere:2003a}
and spin-resonance experiments, which previously suffered from {}``heating''
effects in single dots \cite{vanderwiel:2003a,koppens:2006a}. The
hyperfine interaction mixes the $\ket{S(1,1)}$ and $\ket{T(1,1)}$
states, allowing transport, and effectively removing spin blockade
when these states are nearly degenerate. This behavior leads to a
number of intriguing effects, including stable undriven oscillations
in transport current \cite{ono:2004a,erlingsson:2005b}, and a striking
magnetic-field dependence of leakage current, which allows the extraction
of information about the nuclear spin system \cite{koppens:2005a,jouravlev:2005a}.
Even-odd effects in the spin blockade of many-electron quantum dots
have further revealed the shell-filling illustrated in Figure \ref{cap:DoubleDotGroundState}
\cite{johnson:2005b}.

The influence of spin-dependent terms, causing decoherence or unwanted
evolution, is a central issue in quantum-dot spin quantum computing.
The requirements for fault-tolerant quantum information processing
are very stringent. This raises the bar for required understanding
of these environmental influences to a very high level, and guarantees
that quantum-dot spin decoherence will remain a challenge for some
time to come.

\section{\label{sec:Entanglement}ENTANGLEMENT GENERATION, DISTILLATION, AND
DETECTION}

In addition to the usual requirements for control and coherence, to
demonstrate the true quantum nature of qubits, there have been many
suggestions to create and measure nonlocal multiparticle entanglement
of electron spins in nanostructures \cite{divincenzo:1999b,burkard:2000b,loss:2000a,choi:2000a,egues:2002a,burkard:2003a,samuelsson:2004a,recher:2001a,lesovik:2001a,melin:2001a,costa:2001a,oliver:2002a,bose:2002a,recher:2002a,bena:2002a,saraga:2003a,bouchiat:2003a,recher:2003a,beenakker:2003a,saraga:2004a,egues:2005a}.
These proposals include suggestions to extract spin singlets from
a superconductor through two quantum dots \cite{recher:2001a} or
nanotubes \cite{recher:2002a,bena:2002a}, or to create entanglement
near a magnetic impurity \cite{costa:2001a}, through a single quantum
dot \cite{oliver:2002a}, from biexcitons in double quantum dots \cite{gywat:2002b},
or through a triple dot \cite{saraga:2003a}. It may also be possible
to distill entanglement \cite{bennett:1997a} from an unentangled
Fermi gas through Coulomb scattering in a 2DEG \cite{saraga:2004a}. 

As well as providing a proof of quantum mechanical behavior, entanglement
can be used as a resource for measurement-based quantum computing.
Some measurement-based schemes rely on the creation of highly-entangled
cluster states \cite{raussendorf:2001a}, which could be generated
in quantum-dot arrays using the Heisenberg exchange interaction \cite{borhani:2005c}.
Other measurment-based schemes generate entanglement through partial
Bell-state (parity) measurements \cite{beenakker:2004a}, which could
also be implemented for spins in quantum dots using spin-to-charge
conversion \cite{engel:2005a}. Independent of the method used, the
generation or purification and subsequent detection of entangled electron
spins would present a significant milestone on the road to a working
quantum-dot quantum computer.

\section{\label{sec:Conclusions}CONCLUSIONS AND OUTLOOK}

We have presented some of the theoretical and experimental challenges
to quantum-dot quantum computing with electron spins. The last few
years have seen an extremely rapid rate of progress in experiments
which show that many of the required elements of a spin-based quantum-dot
quantum computer can be realized in principle. The most significant
advances include the reduction of the number of electrons confined
to gated quantum dots down to a single electron \cite{ciorga:2000a},
the demonstration \cite{hanson:2003a} and improvement \cite{hanson:2005a}
of electron spin readout in gated lateral dots, which has led to the
measurement of a spin $T_{1}$ time \cite{elzerman:2004a}, the demonstration
of the $\sqrt{\mbox{{\sc swap}}}$ operation, allowing for the extraction
of an ensemble-averaged $T_{2}^{*}$ time, and spin-echo methods to
extend the decay time within a two-spin encoded subspace \cite{petta:2005a},
and most recently the demonstration of single-spin rotations under
resonant conditions \cite{koppens:2006a}. 

To demonstrate viability of the Loss-DiVincenzo proposal, more experiments
are needed. Although the Loss-DiVincenzo proposal is scalable in principle,
it remains to be seen if there are significant practical obstacles
to scaling-up the number of electrons involved well beyond two.

\subparagraph{Aknowledgments: \textmd{We aknowledge financial support from the
Swiss NSF, the NCCR Nanoscience, EU NoE MAGMANet, DARPA, ARO, ONR,
and JST ICORP.}}

\bibliographystyle{hmmm}
\bibliography{homreview}

\end{document}